\documentclass{aa}

\usepackage{times}
\usepackage{graphics}
\begin{document}

\title{\bfseries 
{\itshape ORFEUS\,} II echelle spectra :\\
H$_2$ measurements in the Magellanic Clouds 
\thanks{Data partly obtained under the
DARA guest observing program in the {\it ORFEUS\,} II Mission}}

\author{
Philipp Richter\inst{1,}\inst{2}
}

\institute{
Sternwarte, Universit\"at Bonn, Auf dem H\"ugel 71, D-53121 Bonn, Germany
\and
Department of Astronomy, University of Wisconsin-Madison, Madison, WI 53706, USA
}

\date{Received 12 November 1999 / Accepted yyy 2000}

\thesaurus{08(03.19.2, 11.09.4, 11.13.1, 13.21.3)}

\offprints{prichter@astro.uni-bonn.de}

\titlerunning{H$_2$ measurements in the Magellanic Clouds}

\maketitle

\begin{abstract}

More than 20 years after the {\it Copernicus} satellite,
{\it ORFEUS} allows the investigation of molecular
hydrogen (H$_2$) in the diffuse interstellar medium by way of FUV
absorption spectroscopy once again.
This time, 
targets in the Magellanic Clouds were also observed,
allowing the first investigation of cold H$_2$ in a metal-poor
environment outside the Milky Way disk.
This paper presents new H$_2$ measurements in LMC gas
along the lines of sight toward HD\,269698 (Sk\,$-$67\,166),
HD\,269546 (Sk\,$-$68\,82) and HD\,36402 (Sk\,$-$67\,104) 
and summarizes the {\it ORFEUS} H$_2$ measurements 
in the 
the Magellanic Clouds.
For seven lines of sight we investigate correlations between 
$N($H\,{\sc i}$)$, $N($H$_{2})$, $E(B-V)$ and compare the 
results with H$_2$ observations in the Milky Way disk. 
We find that the abundance of H$_2$  
is low in comparison to the total gas quantity and speculate
that the
fraction of hydrogen in molecular form is 
limited by the lower dust content of the
Magellanic Clouds. 


\end{abstract}

\section{Introduction}

The molecular hydrogen (H$_2$) is by far the most abundant
molecule in the interstellar medium (ISM) and thus plays a key role
for our understanding of the molecular gas in the ISM
of the Milky Way and other galaxies.
Despite its large abundance, H$_2$ in the ISM is
difficult to measure because it is not seen
in radio emission,
in striking contrast to the second most abundant interstellar molecule,
carbon monoxide (CO). 
H$_2$ emission lines are seen in the near
infrared (NIR), but unfortunately they are weak (quadropole transitions)
and thus can not be used to study the overall interstellar abundance of H$_2$. 
Molecular hydrogen in the diffuse ISM can only be studied by way of
absorption spectrosopy in the far ultraviolet (FUV) toward
stars or other bright UV background sources.
During the seventies,
considerable effort was put into the
investigation of H$_2$ absorption lines 
with the {\it Copernicus} satellite.
Savage et al.\,(1977; hereafter S77) summarized {\it Copernicus}
H$_{2}$ measurements of 102 lines of sight toward nearby
stars in the Milky Way. One of the most striking results was
the correlation between the H$_2$ column
density $N$(H$_2$) and colour excess $E(B-V)$, representative of the
dust amount along a sight line. This relation has been interpreted in terms of the
self-shielding effect of H$_2$ (Federman et al.\,1979).
S77 showed that the transition from low to high molecular
fractions in the local Galactic gas is found at total hydrogen
column densities near
$5.0 \times 10^{20}$\,cm$^{-2}$.

\begin{table*}[t!]
\caption[]{Basic properties of Magellanic Cloud targets}
\begin{tabular}{lllclccccc}
\hline
Object      & Other   & Location & $V$   &    Spectral Type & $E(B-V)$ & $l$ & $b$ & Ref.$^{1}$ & {\it ORFEUS} exp. time \\
Name        & Names   &          & [mag] &                  & [mag]       &     &     &            & [ksec]                 \\
\hline
 LH 10:3120 &  ...         & LMC &  12.80           & O5.5 V    &  $0.17$     & 277.2  & $-$36.1    & 1   & 6.5  \\
 HD\,5980    & Sk\,78       & SMC &  11.80           & OB+WN3    &  $0.07$     & 302.1  & $-$44.9    & 2,3 & 6.8  \\
 HD\,269698 & Sk\,$-$66\,166 & LMC &  12.27           & O4\,If    &  $0.09$     & 277.8  & $-$32.5    & 4,5 & 6.2  \\
 HD\,269546  & Sk\,$-$68\,82  & LMC &  11.30           & B3\,I+WN3 &  $\le 0.02$ & 279.3  & $-$32.8    & 4,6 & 6.4  \\
 HD\,36402   & Sk\,$-$67\,104 & LMC &  11.50           & OB+WC5    &  $\le 0.02$ & 277.8  & $-$33.0    & 2,7 & 7.8  \\
\hline
\end{tabular}
\noindent

$^{1}$ References: (1) Parker et al. (1992); (2) Sembach \& Savage (1992); (3) Fitzpatrick \& Savage (1983);
(4) Chu et al. (1994); \\
(5) Wilcots et al. (1996); (6) Vacca \& Torres-Dodgen (1990); (7) de\,Boer \& Nash (1982) and references therein
\end{table*}

Measurements with {\it Copernicus}, however,
were limited to the very local interstellar gas of the Milky
Way. For more distant background sources, {\it Copernicus}
was not sensitive enough. Later UV satellites, such as 
{\it IUE} and {\it HST}, had better sensitivity,
but these instruments do not
cover the wavelength range below 1150 \AA\, where
the transitions of H$_2$ are seen.
The analysis of extragalactic H$_2$ gas is of great 
importance since the abundance of H$_2$ can be
studied in environments very different from those of 
the Milky Way. 
The Magellanic Clouds, the most nearby satellite
galaxies of the Milky Way, are ideal hunting grounds for
extragalactic H$_2$ measurements, because
they provide many bright stars suitable as 
UV background sources for absorption spectroscopy.
Moreover, H$_2$ has been detected from warm regions
in both galaxies in the near-IR emission lines 
(Koornneef \& Israel 1985,
Israel \& Koornneef 1988, 1991).
It has been suggested that, due to the
lower metallicity and the lower dust content,
the amount of H$_2$ in the diffuse interstellar medium 
of the Magellanic Clouds is
significantly lower than in the Milky Way (Clayton et al. 1995).

The {\it ORFEUS} telescope, launched for its second mission
in 1996, was the first instrument able to measure H$_2$
absorption lines in the LMC (de Boer et al. 1998) and 
SMC (Richter et al. 1998). In addition, the spectrum of 
LH 10:3120 was used to detemine an upper 
limit for the H$_2$/CO ratio in the LMC gas along this
line of sight (Richter et al. 1999a).
Together with the observations presented here, these {\it ORFEUS} spectra provide
the first opportunity to investigate the relations between 
$N($H\,{\sc i}$)$, $N($H$_{2})$ and $E(B-V)$ in diffuse interstellar gas of the
Magellanic Clouds in comparison to the Milky Way.

\section{Observations}

The observations have been carried out 
during the second mission of {\it ORFEUS} on the 
{\it ASTRO-SPAS} space shuttle mission in Nov./Dec. 1996.
{\it ORFEUS} is equipped with two alternatively
operating spectrometers, the echelle spectrometer 
(Kr\"amer et al. 1990) and the Berkely spectrometer
(Hurwitz \& Bowyer 1996).
The spectroscopic data presented here
were obtained with the Heidelberg/T\"ubingen 
echelle spectrometer. This instrument has a
resolution of somewhat better than $\lambda / \Delta \lambda = 10^4$
(Barnstedt et al. 1999), working in the spectral range between
$912$ and $1410$ \AA.
A detailed description of the instrument is given by
Barnstedt et al.\,(1999).

Here we study the {\it ORFEUS} spectra of 4 LMC stars and one SMC star.
Basic 
information about the targets is given in Table 1.
The primary data reduction was performed by the 
{\it ORFEUS} team in T\"ubingen (Barnstedt et al.\,1999). 
In order to improve signal-to-noise ratios (S/N),
all spectra have been filtered by a wavelet algorithm
(Fligge \& Solanki 1997). The resolution after
filtering is $\sim 30$ km\,s$^{-1}$.
Heliocentric velocities have been transformed 
for each line of sight into
the LSR (Local Standard of Rest) system.

\section{Data analysis}

The complex line-of-sight structure in direction of the Magellanic Clouds,
with contributions from local Galactic gas (0 km\,s$^{-1}$),
Galactic halo gas (near $+60$ and $+120$ km\,s$^{-1}$ in front
of the LMC) and
Magellanic Cloud gas (near $+250$ km\,s$^{-1}$ for the LMC
and $+150$ km\,s$^{-1}$ for the SMC;
see Savage \& de\,Boer 1979, 1981; de Boer et al. 1980; Bomans
et al. 1996) makes the thorough analysis of H$_2$ absorption
lines at LMC velocities and $\sim 30$ km\,s$^{-1}$ resolution
a difficult task. The main problem is that for
the vast majority of the H$_2$
transitions line blends from atomic or molecular species
can not be excluded, even when many of these blendings
might are unlikely. As a consequence, the number of unambigously
identified H$_2$ features at high radial velocities
is strongly limited to only a few wavelength regions.
Typical line strengths
for low H$_2$ column densities (as observed in the
spectra presented in the following) have values $\le 100$\,m\AA,
which is (at low S/N) comparable with the strength
of noise peaks.
For most of these lines, only
upper limits for the equivalent widths can be obtained.
For diffuse H$_2$ clouds it is known that
the process of UV pumping (Spitzer \& Zweibel 1974) 
often leads to an excitation of the higher rotational states,
particulary if the total H$_2$ column density is below the limit
for the self-shielding effect. Thus, the rotational
excitation of H$_2$ in the most diffuse medium often does 
{\it not} reflect the actual kinetic temperature of the gas. 
H$_2$ line strengths in diffuse clouds
for excited rotational states might be
significantly higher than for the ground states
(see the {\it Copernicus} spectrum of
$\zeta$ Pup, as published by  Morton \& Dinerstein 1976, where the
strongest H$_2$ lines occur for $J=3$),
even if the gas is cold.
In the most complex case, the equivalent
widths for the ground state lines are below the
detection limit, while in the same spectrum, absorption from
higher rotational levels is visible. 

Velocity information 
from metal lines as well as model spectra
for the excited rotational states were used to interpret 
the complex H$_2$ absorption pattern
found in the {\it ORFEUS} spectra of stars
in the Magellanic Clouds.

\begin{figure}[h!]
\resizebox{0.81\hsize}{!}{\includegraphics{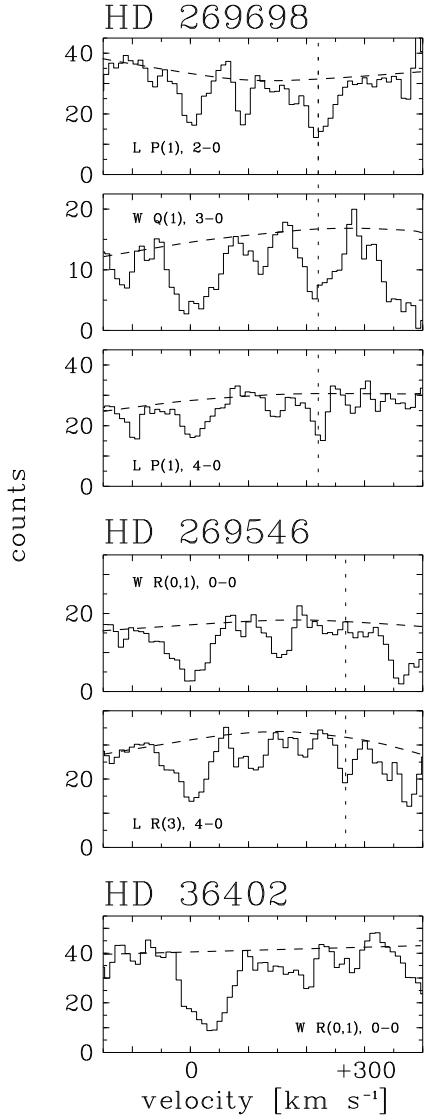}}
\caption[]{
Interstellar H$_2$ line profiles for HD\,269698,
HD\,269546 and HD\,36402 are shown, plotted in counts versus
LSR velocity units.
For HD\,269698, H$_2$ absorption from LMC gas is
clearly visible near $+220$ km\,s$^{-1}$. For HD\,269546,
weak H$_2$ absorption might be present 
near $+265$ km\,s$^{-1}$, but the origin of
the absorption feature in the R(3), 4-0 line
remains doubtful.
No H$_2$ absorption
at LMC velocities is seen in the {\it ORFEUS}
spectrum of HD\,36402.
The LMC velocities found from the H$_2$ line
centres have been marked with dashed lines.
In all three spectra,
H$_2$ absorption from Galactic gas is
present near 0 km\,s$^{-1}$.
The adopted continuum is indicated for each of
the profiles
}
\end{figure}

\section{\bfseries{\itshape ORFEUS} H$_2$ measurements}

For the lines of sight toward HD\,269698, HD\,269546 and HD\,36402 the
analysis of H$_2$ line strengths is presented in the
following.
For 2 other lines of sight, {\it ORFEUS} H$_2$ measurements of 
Magellanic-Cloud gas have recently been published by de Boer et al.\,
(1998; LH\,10:3120) and
Richter et al.\,(1998, 1999a; HD 5980, LH\,10:3120).

Wavelengths and oscillator strengths for the
H$_2$ lines have been taken from the list
of Morton \& Dinerstein (1976)
\footnote{We note that the Lyman P(1), 10-0
line is located at $982.839$ \AA\, and
not at $982.383$ \AA, as given by Morton \&
Dinerstein (1976). See the wavelength list 
of Abgrall \& Roueff (1989).}.
We measured equivalent widths ($W_{\lambda}$) by using either trapezoidal
or gaussian fits. For the error determination we
used the algorithm of Jenkins et al. (1973), taking 
into account photon statistics and the number of
pixels involved for each line. 
In order to estimate the uncertainty for the 
choice of the continuum,
we fitted a maximum
and a minimum continuum to the data
in the vicinity of each line and derived
a mean deviation. The error for
$W_{\lambda}$ given in Table 2
represents the total uncertainty calculated
from all contributions discussed above.
Column densities were
derived by using a standard curve-of-growth technique.

\subsection{HD\,269698}

The {\it ORFEUS} spectrum of HD\,269698 in the
Large Magellanic Cloud 
shows weak H$_2$ absoption at LMC velocities near 
$+220$ km\,s$^{-1}$. Six lines from the two
rotational ground states ($J=0,1$)
with high oscillator strengths are clearly seen in the spectrum
and are not blended by other transitions. For additional
five lines from higher rotational states we find
upper limits for the equivalent widths 
of $W_{\lambda} \le 82$\,m\AA\ (Table 2). 
Fig.\,1 shows three of the detected H$_2$ absorption lines plotted on a
velocity scale. 
The lack of absorption in higher 
rotational states indicates that the H$_2$ gas is not strongly excited.
Constructing curves of growth for each
rotational state we obtain
column densities of $4.0^{+4.0}_{-2.0} \times 10^{14}$ cm$^{-2}$
for $J=0$ and $5.0^{+6.0}_{-3.0} \times 10^{15}$ cm$^{-2}$ for the
$J=1$ state, using a $b$ value of 8 km\,s$^{-1}$ (best fit).
The total H$_2$ column density in the LMC gas 
toward HD\,269698, derived by summing
over $N(0)$ and $N(1)$, 
is $N$(H$_2$)$_{\rm total} = 
5.4^{+7.3}_{-3.1}  \times 10^{15}$ cm$^{-2}$.
The error is derived from the uncertainty
for the fit to the curve of growth and includes the error
for the individual equivalent widths and the uncertainty for
the $b$ value. From the detection limits for 
the lines from $J \ge 2$ we can exclude the possibilty that
the higher rotational states will significantly contribute to the
total H$_2$ column density. 

HD\,269698 is located in the OB
association N\,57 at the rim
of the supergiant shell LMC\,4 where the H\,{\sc i}
emission (Rohlfs et al. 1984) has a minimum.
The {\it IUE} spectrum of HD\,269698 
(Domg\"orgen et al. 1994) reveals three S\,{\sc ii} components
at LMC velocities in front of the star:
near $+220$ km\,s$^{-1}$, near $+245$ km\,s$^{-1}$
and near $+290$ km\,s$^{-1}$. The detected H$_2$ 
lines obviously belong to the first component.

\subsection{HD\,269546}

In the {\it ORFEUS} spectrum of the LMC star HD\,269546,
no clear H$_2$ absorption is visible at LMC velocities.
The presence of H$_2$ at LMC velocities (near 
$+200$ km\,s$^{-1}$) in {\it ORFEUS} data
was indicated by
Widmann et al. (1998), using 
a coaddition of 25 Lyman- and Werner lines.
However, the Werner R(0), R(1) line-pair
(in Fig.\,1 plotted on a
velocity scale) gives no hint for
the presence of H$_2$ absorption at LMC velocities.
Marginal H$_2$ absorption might be present in the
Lyman P(1), 2-0 line ($\lambda = 1078.925$ \AA)
and in the Lyman R(3), 4-0 line ($\lambda = 1053.976$ \AA)
near $+265$ km\,s$^{-1}$ (Fig.\,1), but
these absorption features are not clearly 
distinguishable from noise peaks and
no other H$_2$ profiles from $J=1,3$ show similar features
at $+265$ km\,s$^{-1}$.
Metal lines in the LMC gas near $+250$ km\,s$^{-1}$ have been found
in the {\it IUE} spectrum of HD\,269546 (Grewing \& Schultz-Luepertz 1980).
Moreover, the {\it IUE} data reveal 
absorption over the whole
velocity range between $0$ and $+290$ km\,s$^{-1}$,
most likely related to Galactic halo gas and weaker 
LMC components. H$_2$ absorption at $+120$ km\,s$^{-1}$
is seen in some of the stronger lines, indicating 
that the Galactic high-velocity gas in front of HD\,269546
contains molecular gas and dust
(Richter et al.\,1999b).
The H\,{\sc i} emission line
data from Rohlfs et al.\,(1984) show the
LMC gas at $+250$ km\,s$^{-1}$.
HD\,269546 is member of the OB association
LH\,58 in the N\,144 superbubble
complex northwest of 30\,Doradus.
The H\,{\sc i} gas seen in 21\,cm emission at $+250$ km\,s$^{-1}$ 
is most likely in front of N\,144.

Detection limits for eight H$_2$ absorption lines
at LMC velocities near $+265$ km\,s$^{-1}$ are used to obtain upper limits for the column 
densities of $N(J)$ for $J \le 4$ by fitting the values of log ($W_{\lambda}/\lambda$)
to the linear part of the curve of growth.
We calculate an upper limit for the total
H$_2$ column density by modeling the population
of the rotational states for T$_{\rm ex} \leq 300$ K.
From that we derive 
$N$(H$_2$)$_{\rm total} \le 2.3 \times 10^{15}$ cm$^{-2}$
for the LMC gas toward HD\,269546.

\begin{table}[t!]
\caption[]{H$_2$ equivalent widths for LMC gas toward HD\,269698}
\begin{tabular}{lllcccc}
\hline
\hline
\rule [-2mm]{0mm}{6mm}{Line} & & $\lambda$ [\AA] & $f$  & $W$ [m\AA] \\
\hline
\hline
\multicolumn{5}{c}{\rule [-2mm]{0mm}{6mm}{$J=0$, $g_J=1$, $E_J=0$ eV}}\\
\hline
\noalign{\smallskip}
W & R(0,1),0-0 & $1008.553$ & $0.04480$   & present   \\
L & R(0),2-0 & $1077.138$ & $0.01190$ & $45 \pm 14$ \\
L & R(0),4-0 & $1049.366$ & $0.02350$ & $\le 55$     \\
\noalign{\smallskip}
\hline
\multicolumn{5}{c}{\rule [-2mm]{0mm}{6mm}{$J=1$, $g_J=9$, $E_J=0.01469$ eV}}\\
\hline
\noalign{\smallskip}
L & P(1),2-0 &  $1078.925$ & $0.00385$ & $82 \pm  19$ \\
L & P(1),3-0 &  $1064.606$ & $0.00584$ & $\le 82$     \\
L & R(1),4-0 &  $1049.958$ & $0.01600$ & $88 \pm  25$ \\
L & P(1),4-0 &  $1051.031$ & $0.00749$ & $77 \pm  25$ \\
W & Q(1),3-0 &   $947.425$ & $0.02730$ & $111 \pm 26$\\
\noalign{\smallskip}
\hline
\multicolumn{5}{c}{\rule [-2mm]{0mm}{6mm}{$J=2$, $g_J=5$, $E_J=0.04394$ eV}}\\
\hline
\noalign{\smallskip}
L & R(2),4-0 & $1051.497$ & $0.01470$ & $\le 34$ \\
W & Q(2),0-0 & $1010.941$ & $0.02380$ & $\le 59$ \\
\noalign{\smallskip}
\hline
\multicolumn{5}{c}{\rule [-2mm]{0mm}{6mm}{$J=3$, $g_J=21$, $E_J=0.08747$ eV}}\\
\hline
\noalign{\smallskip}
L & R(3),4-0 &  $1053.976$ & $0.01420$ & $\le 47$  \\
\noalign{\smallskip}
\hline
\end{tabular}
\noindent
\end{table}

\subsection{HD\,36402}

No H$_2$
absorption is
seen in the spectrum of HD\,36402
at LMC velocities in the range
$+220$ to $+320$ km\,s$^{-1}$.
In this velocity range, atomic absorption has been
found by de Boer \& Nash (1982).
We place an upper limit on
the H$_2$ column density in the LMC gas after inspecting 
some of the strongest of the H$_2$ transitions in the rotational
states $J=0$ to $J=4$.
HD\,36402 is located in the superbubble N\,51D and
shows hydrogen emission and metal absorption from
LMC foreground gas near $+300$ km\,s$^{-1}$
(de Boer \& Nash 1982). Therefore we expect 
H$_2$ absorption from LMC gas roughly at 
the same velocity. Inspecting the R(0), R(1)
line-pair plotted on the velocity scale (Fig.\,1,
lowest panel) there is very weak absorption 
near $+290$ km\,s$^{-1}$, but this feature
has no significance with respect to the noise.

In the same way as for HD\,269546, we determine
upper limits for the individual column densities 
$N(J)$ for $J \leq 4$ from the detection limits 
for some of the stronger H$_2$ transitions 
near $+300$ km\,s$^{-1}$.
We find an upper limit for the total
H$_2$ column density in the LMC gas toward HD\,36402
of $N$(H$_2$)$_{\rm total} \le 1.0 \times 10^{15}$ cm$^{-2}$.
Again, this upper limit is calculated under the assumption that the excitation
temperature of possibly existing H$_2$ gas in the LMC
would
not exceed a value of 300 K.

\section{H\,{\sc i} measurements}

For two of our sight lines (HD\,269698 and HD\,269546) we present
the determination of 
H\,{\sc i} column densities from the analysis of the  
Ly\,$\alpha$ absorption near 1215 \AA. 
The values for the H\,{\sc i} column densities along the other
three lines of sight have been adopted from
the literature. 
All H\,{\sc i} column densities 
are summarized in Table 3.
For HD\,5980 and HD\,36402, the column density of H\,{\sc i} has
been determined by Fitzpatrick \& Savage (1983) and 
de Boer \& Nash (1982), respectively, using H\,{\sc i} 
emission line data in combination
with the Ly\,$\alpha$ absorption near 1215 \AA.
They derive H\,{\sc i} column densities of $N$(H\,{\sc i}$)
= 1.0 \times 10^{21}$ cm$^{-2}$ for the SMC gas toward HD\,5980 and
$1.6 \times 10^{20}$ cm$^{-2}$ for the LMC gas toward 
HD\,36402. 
For LH\,10:3120, the LMC H\,{\sc i} column density
is $2.0 \times 10^{21}$ cm$^{-2}$, 
obtained by a multi-component fit 
of the Ly\,$\alpha$ profile
(Richter et al.\,1999a).

We use a similar technique for the determination
of H\,{\sc i} column densities in the LMC gas 
toward HD\,269698 and
HD\,269546. 
For HD\,269546, 
we fix the LMC component 
at a velocity of $+265$ km\,s$^{-1}$, similar to
the velocity for which we had determined the upper limit 
for the H$_2$ column density in Sect.\,4.2 . The 
velocity structure seen in metal lines (Grewing \& Schulz-Luepertz 1980), however,
indicates that there are definitely additional (weaker) 
absorption components in front
of HD\,269546. We thus might slightly overestimate the
H\,{\sc i} column density in the LMC gas at $+265$ km\,s$^{-1}$
by fitting one single LMC component to
the Ly\,$\alpha$ absorption structure.
The situation is even more difficult for the Ly\,$\alpha$ profile
in the spectrum of HD\,269698. {\it IUE} data
of HD\,269698 show the presence of three 
velocity components in this sight line (Domg\"orgen et al.\,1994),
near $+220, +245$ and $+290$ km\,s$^{-1}$.
The H\,{\sc i} emission (Rohlfs et al.\,1984) shows
a weak component near $+256$ km\,s$^{-1}$ which 
could be associated with the absorption component
near $+245$ km\,s$^{-1}$ (see
Domg\"orgen et al.\,1994). The S\,{\sc ii} 
abundances found by Domg\"orgen et al. indicate
similar total gas quantities for the two
main components at $+245$ km\,s$^{-1}$ and
at $+220$ km\,s$^{-1}$. The H$_2$ absorption was
found in the latter component (see Sect. 4.1).
For the Ly\,$\alpha$ fit, we fix the two LMC
components at $+220$ (cloud I) and $+245$ (cloud II) km\,s$^{-1}$,
assuming equal H\,{\sc i} column densities. 
For the fitting procedure we use a multi-component
Voigt profile.
Galactic foreground absorption by H\,{\sc i}
is taken into account by a fit component
at $0$ km\,s$^{-1}$.
The multiple
velocity components are not resolved in the Ly\,$\alpha$
profile and we do not take into account additional
absorption from Galactic intermediate and high-velocity gas.
Thus, it is clear that our results derived by this method represent
only rough estimates for the distribution of the H\,{\sc i}
gas in front of the stars. However, for the
comparison between $N($H\,{\sc i}$)$, $N($H$_{2})$ and $E(B-V)$, as presented
in Sect.\,7, the determined H\,{\sc i} column densities
are sufficiently accurate.

\begin{figure*}
\resizebox{12cm}{!}{\includegraphics{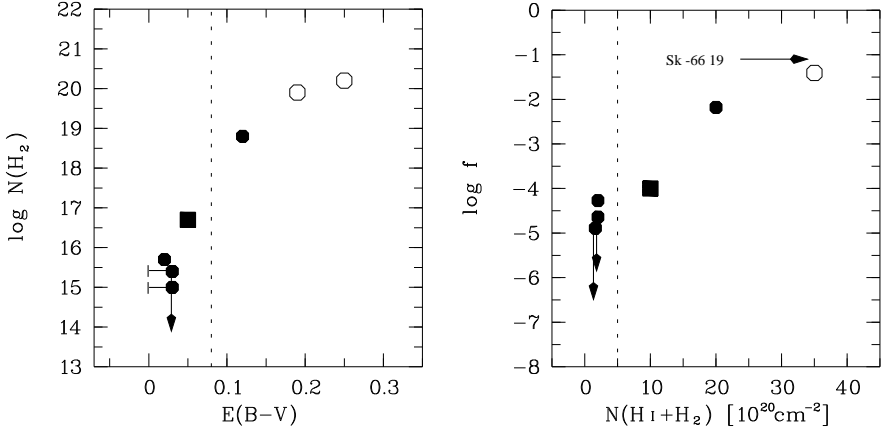}}
\hfill
\parbox[b]{40mm}{
\caption[]{
Correlations between atomic hydrogen, molecular hydrogen and colour excess
for Magellanic Cloud gas along seven lines of sight. LMC stars measured with
{\it ORFEUS} are labeled by filled dots, the one SMC target is given by a filled square.
Data from Gunderson et al. (1998) for two additional lines of sight to the LMC
have been included, here given as open circles. The plots are discussed
in Sect.\,7
}}
\end{figure*}

\begin{table*}
\caption[]{LMC gas properties along seven lines of sight}
\begin{tabular}{lcccccrc}
\hline
Object & Location & $E(B-V)_{\rm LMC}^1$ & log $N$(H$_2$) & log $N$(H\,{\sc i})
& $N$(H\,{\sc i} + H$_2$)/$E(B-V)$ & log $f$ &  Ref.$^{2}$ \\
       &          & [mag] &   & & [cm$^{-2}$]  &     \\
\hline
LH 10:3120    & LMC & $0.12 \pm 0.03$       & $18.8 \pm 0.3^1$ & $21.3^{+0.2}_{-0.4}$ & $1.7 \times 10^{22}$      & $-2.18$     & 1,2\\
HD 5980       & SMC & $0.05 \pm 0.02$       & $16.7 \pm 0.4^1$ & $21.0 \pm 0.2$       & $2.0 \times 10^{22}$      & $-4.00$     & 3,4\\
HD\,269698    & LMC & $0.02^{+0.03}_{-0.02}$ & $15.7 \pm 0.4^1$ & $20.3 \pm 0.1$       & $1.0 \times 10^{22}$      & $-4.27$     & 5\\
HD\,269546    & LMC & $\leq 0.03$           & $\le 15.4$       & $20.3 \pm 0.1$       & $\geq 6.7 \times 10^{21}$ & $\le -4.64$ & 5\\
HD\,36402     & LMC & $\leq 0.03$           & $\le 15.0$       & $20.2^{+0.1}_{-0.2}$ & $\geq 5.3 \times 10^{21}$ & $\le -4.89$ & 5,6\\
Sk $-$66\,19  & LMC & $0.25 \pm 0.03$       & $20.2 \pm 0.1^3$ & $21.9^{+0.2}_{-0.3}$ & $3.0 \times 10^{22}$      & $-1.37$ & 7,8 \\
Sk $-$69\,270 & LMC & $0.19 \pm 0.03$       & $19.9 \pm 0.2^3$ & $21.5^{+0.2}_{-0.3}$ & $1.9 \times 10^{22}$      & $-1.41$ & 7,8 \\
\hline
\end{tabular}

\noindent
$^{1}$ Errors include uncertainty for the choice of the doppler parameter $b$\\
$^{2}$ References: (1) de Boer et al.\,(1998); (2) Richter et al.\,(1999a);
(3) Richter et al.\,(1998); (4) Fitzpatrick \& Savage (1983); \\
(5) this paper; (6) de\,Boer \& Nash (1982); (7) Gunderson et al.\,(1998); (8) Fitzpatrick (1985)\\
$^{3}$ From intermediate resolution spectra with {\it HUT}\\
\end{table*}

The Ly\,$\alpha$ fit for HD\,269546 provides the best agreement with the data
with a Galactic foreground absorption of
$N($H\,{\sc i}$)_{\rm MW} = 3.5 \pm 0.6 \times 10^{20}$ cm$^{-2}$ 
and an additional LMC component at $+265$ km\,s$^{-1}$ of 
$N($H\,{\sc i}$)_{\rm LMC} = 2.0 \pm 0.5 \times 10^{20}$ cm$^{-2}$.
For HD\,269698, the best fit is found for
$N($H\,{\sc i}$)_{\rm MW} = 3.5 \pm 0.7 \times 10^{20}$ cm$^{-2}$
and $N($H\,{\sc i}$)_{\rm LMC, comp.\,I} \equiv
N($H\,{\sc i}$)_{\rm LMC, comp.\,II} = 1.0 \pm 0.4 \times 10^{20}$ cm$^{-2}$
for the two LMC clouds at $+220$ and $+245$ km\,s$^{-1}$.
The H\,{\sc i} column densities in the LMC gas in
these two lines of sight are significantly 
lower than found for HD\,5980 and LH\,10:3120, but
comparable with the H\,{\sc i} column density found in the LMC gas
toward HD\,36402.

\section{Colour excess $E(B-V)$}

The colour excess $E(B-V)$ for each line of sight, adopted from
different publications, is given in Table 1. 
All lines of sight show total values of $E(B-V)$ lower 
than 0.20\,mag. 
The main problem is to separate the contributions
of the Galactic foreground from the colour excess within
the Magellanic Clouds. 
For LH\,10:3120\,(LMC) and HD\,5980\,(SMC), we adopt
values for $E(B-V)_{\rm LMC}$ from previous publications
(de Boer et al. 1998; Richter et al. 1998).
For HD\,269698 and HD\,269546, we have calculated
the foreground extinction from the H\,{\sc i} 
column densities of the Galactic foreground gas by
using the mean gas-to-colour excess relation, as given
by Bohlin et al. (1978).
The values for $E(B-V)_{\rm LMC}$ 
are presented in Table 3. 
Their errors are based on the uncertainty for 
$E(B-V)_{\rm total}$ (as cited in the literature;
see Table 1) in addition to the uncertainty 
for the Galactic foreground reddening, which shows variations in the
range of $0.05$\,mag in direction of the LMC and
$0.02$\,mag toward the SMC (Bessel 1991).

\section{Discussion}

Values of $N($H\,{\sc i}$)$, $N($H${_2})$ and $E(B-V)$ for all
five lines of sight measured with {\it ORFEUS} (summarized in Table 3) are used 
to investigate the diffuse molecular ISM of the Magellanic Clouds.
In order to extend our data, we include recent results from Gunderson et al. (1998)
for two lines of sight toward Sk $-$66\,19 and Sk $-$69\,270 in the LMC.
They estimate, on the basis of low dispersion spectra
with the Hopkins Ultraviolet Telescope ({\it HUT}), column densities of 
molecular hydrogen in the LMC gas by fitting H$_2$ line profiles (see Table 3).

Fig.\,2 presents correlations 
between $N($H$_2)$, $E(B-V)$, $f= 2N($H$_2$$)/[N($H\,{\sc i})$ + 2N($H$_2)]$, 
and $N($H\,{\sc i}$ + $H$_2) = N($H\,{\sc i}$) + 2N($H$_2)$.
The left panel shows log $N($H$_2)$ plotted versus $E(B-V)$.
In principle, we find the typical relation known from the 
{\it Copernicus} H$_2$ survey from S77 for Galactic gas.
In both Milky Way and the Magellanic Clouds the
logarithmic H$_2$ column density (log $N($H$_2)$) 
undergoes a transition from low to to high values 
at $E(B-V) \approx 0.08$ (dashed line) due to the
self-shielding effect of H$_2$ 
(Federman et al.\,1979).

It is known that the Magellanic Clouds have
a significantly lower dust content than the Milky Way.
Typical gas-to-dust ratios
$N($H\,{\sc i}$ + $H$_2)$/$E(B-V)$ in the Magellanic Clouds
are 4 times (LMC) and 8 times (SMC) higher than in Milky Way gas
(Koornneef 1982; Bouchet et al.\,1985, respectively).
In our sample, we find gas-to-dust ratios
as high as $3.0 \times 10^{22}$\,cm$^{-2}$
for the gas in the LMC and SMC (see Table 3), consistent with
these results. For HD\,36402, HD\,269698 and HD\,269546,
the ratios are significantly lower, but note that
these values most likely represent lower limits due to the
large uncertainty for the gas-to-dust ratio near the
zero point of the $E(B-V)$ scale. 
With respect to the 
generally lower
dust content
and the
relation between H$_2$ column density
and $E(B-V)$ (Fig.\,2, left panel) 
one should expect
that the fraction of gas in molecular form is
significantly lower in the Magellanic Cloud than
in the Milky Way. 
From more theoretical considerations,
Elmegreen (1989) concluded that interstellar clouds in 
Magellanic type irregular galaxies should be
mostly atomic, since their lower metallicity   
directly influences the shielding function $S$ for
the cloud layers. This author also showed that 
the H to H$_2$ conversion also depends sensitively 
on the pressure and radiation field in the ISM (Elmegreen 1993). 
Accordingly, typical sight lines through the Magellanic
Clouds might not contain any measureable column density
of H$_2$, except for those, whose column density
in H\,{\sc i} is high enough to allow a significant fraction of 
the gas to convert into molecular form.

As the right panel of Fig.\,2 shows, the discussed effects are
slightly visible in the FUV absorption line data.
The figure shows the molecular fraction $f$ plotted against the total hydrogen column
density $N($H\,{\sc i}$ + $H$_2)$.
The {\it Copernicus} sample (S77) shows that 
the transition from low ($f \le 10^{-2}$) to high ($f > 10^{-2}$) 
molecular fractions in the local Galactic gas is found
at a total hydrogen column density 
(`transition column density' $N_{\rm T}$(H\,{\sc i}))
near $5.0 \times 10^{20}$ cm$^{-2}$ (right panel, dashed line).
We find 
high total hydrogen column densities
($\ge 10^{21}$ cm$^{-2}$) but low molecular hydrogen fractions ($f \le 10^{-2}$)
for the Magellanic Clouds gas along
two of seven lines of sight.
The data points of these two lines of sight toward LH\,10:3120 and 
HD\,5980 indicate that the transition column density $N_{\rm T}$
from low to high molecular fractions 
could be indeed higher in the Magellanic Clouds than
in the Milky Way. Only for sight lines with
a very high total hydrogen column density (Sk $-66$\,$19$ and
Sk $-69$\,$270$), the molecular fractions exceeds values above
1 percent. For sight lines with $N($H$_{\rm total}) 
\leq 10^{21}$ cm$^{-2}$ the molecular fractions in the 
Magellanic Cloud gas seem to be negligible.

Additional sight line measurements
toward the Magellanic Clouds, however,
are required to investigate these relations on a 
statistically more significant level.  
With a larger data set it might be 
possible to determine the transition
column density from low to high molecular fractions as
a function of the overall metallicity.
Since it is known that the SMC is even more metal-poor than
the LMC, it is of special interest to also investigate 
differences in the molecular gas fractions between
LMC and SMC.
For that, the {\it FUSE} satellite, launched in June 1999,
holds the prospect for fresh H$_2$ absorption
line data in the near future.

\acknowledgements
I thank K.S. de\,Boer and the Heidelberg-T\"ubingen team
for permission to use the G.O. and P.I. data on
Magellanic Cloud star spectra for this study and for
their great support.
I thank K.S. de\,Boer for helpful comments on this work.
When this paper was prepared, PR was supported by a grant from 
the DARA (now DLR) under code 50 QV 9701 3.

{}
\end{document}